\documentstyle[prl,aps,preprint]{revtex}
\begin{document}
\draft

\renewcommand{\baselinestretch}{1.5}

\title{Three-body halos.\\ III. Effects of finite core-spin}
\author{D.V.~Fedorov\thanks{On leave from the Kurchatov Institute,
 123182 Moscow, Russia} \\ Institute of Physics and Astronomy, \\
Aarhus University, DK-8000 Aarhus C \\ and \\ European Centre for Theoretical
Studies in Nuclear Physics \\ and Related
Areas,  I-38050 Trento, Italy \\ and \\
 E.~Garrido\thanks{Supported by Human Capital and Mobility contract
 nr. ERBCHBGCT930320} and A.S.~Jensen \\
Institute of Physics and Astronomy, \\
Aarhus University, DK-8000 Aarhus C}

\maketitle

\begin{abstract}
We investigate effects of finite core-spin in two-neutron halos in the
three-body model consisting of two neutrons and a core. The states in
the neutron-core subsystem can then have an additional splitting due
to the coupling of the neutron and core spins. The connection between
the structure of the total system and the states of the neutron-core
subsystem depends strongly on this spin splitting. However, the
spatial structure is essentially unchanged. The neutron momentum
distribution in a nuclear dissociation reaction will consist of a
broad and a narrow distribution arising from the two spin split
neutron-core virtual states. The large distance behavior is
investigated and the conditions for the Efimov effect turn out to be
more restrictive. The magnetic moment is equal to the core-value
independent of spin splitting and energies of the neutron-core states.
The excited states of opposite parity depend strongly on the
spin splitting.  The parameters in our numerical examples are chosen
to be appropriate for $^{11}$Li and $^{19}$B.
\\PACS numbers 21.45.+v, 21.60.Gx
\end{abstract}


\section{Introduction}
The existence of halo nuclei is by now established
\cite{han87,han93,tan91} as weakly bound and spatially extended systems.
A rather accurate description is obtained by use of two- and
three-body models.  Criteria for the occurrence of halos and the
related asymptotic large distance behavior were recently discussed
\cite{rii92,fed93a,fed94a,fed94b}.  These investigations assumed many
simplifying approximations, which allowed extraction of general
properties and therefore provided insight into the general nature of
such systems.  Of particular interest are bound three-body systems
where furthermore all two-particle subsystems are unbound. These
so-called Borromean systems \cite{zhu93} appear for a substantial
range of two-body potentials \cite{ric94}.

The nucleus $^{11}$Li has played an especially dominating role in the
developments. The approximate description as a three-body system
consisting of two neutrons outside the $^9$Li-core was only recently
suggested \cite{joh90,ber91,zhu91}. Since then a series of improved
calculations appeared, see for example \cite{zhu93}. The most detailed
and accurate of these three-body computations are based on the Faddeev
equations \cite{tho93}. In spite of the large number of publications
the possibly significant consequences of the finite core-spin $s_c$
has so far not been considered at all in three-body calculations.  The
neutron- and the core spins can couple to a total angular momentum of
$s_c \pm \frac{1}{2}$, while the neutron-core motion still remains in
the same relative orbital state.  If the interaction is independent of
these couplings, the structure corresponds to that of a spin-zero
core.  This is, however, highly improbable considering the well known
appreciable spin dependence both of the strong interaction and of the
observed bound states and resonances throughout the periodic table.
The states in neighboring nuclei like $^{12}$B, $^{12}$N and $^{8}$Li
as well as computations are in particular suggestive for a similar
spin splitting in $^{10}$Li, see ref.~\cite{ajz88,boh93}.

The purpose of this paper is to investigate the consequences of finite
core-spin for halo nuclei.  The resulting spin splitting is probably
an important ingredient for understanding the relation between the
ground state structures of the neutron-core subsystem and the total
three-body system, e.g~$^{10}$Li and $^{11}$Li.  Another effect is
that the peculiar Efimov structure \cite{fed93b}, which might show up
in halo nuclei, is hindered by the spin splitting of the virtual
s-states. As in our earlier papers \cite{rii92,fed94a,fed94b}, we want
to establish the general features, qualitatively or quantitatively as
best we can, and apply the results on realistic examples. We shall
again for convenience assume that all three particles are inert and
all the core degrees of freedom are therefore frozen. We shall
furthermore assume throughout the paper that the total angular
momentum $J$ for the ground state is equal to $s_c$ and that the total
parity equals that of the core.

The paper is the third (after \cite{fed94a,fed94b}) in a series of
papers discussing various aspects of three-body halos.  After the
introduction we sketch in section 2 the general theoretical method
used to compute the properties of the three-body system. The
interactions between the different two-body subsystems are
parametrized in section 3. In section 4 the asymptotic form of the
Faddeev equations is derived for interactions causing the spin
splitting and in section 5 is discussed the large distance behavior in
connection with the Efimov effect.  The resulting spatial structure
and the measured momentum distributions are discussed in section 6.
Magnetic moment is investigated in section 7 and excited states of
$1^-$ character on top of the ground state in section 8.  Finally we
give a summary and the conclusions in section 9.

\section{Method}
The Hamiltonian of the system, where the center of mass kinetic energy
is subtracted, is given by
\begin{equation} \label{e1}
	H =\sum_{i=1}^{3} \frac{p_{i}^2}{2m_{i}} - \frac{P^2}{2M} +
        \sum_{i>j=1}^{3} V_{ij} \; ,
\end{equation}
where $m_{i}$, ${\bf r}_i$ and ${\bf p}_i$ are mass, coordinate and
momentum of the $i$'th particle, $V_{ij}$ are the two-body potentials,
$P$ and $M$ are respectively the total momentum and the total mass of
the system. We shall use the Jacobi coordinates basically defined as
the relative coordinates between two of the particles (${\bf x}$) and
between their center of mass and the third particle (${\bf y}$).  The
precise definitions and the corresponding three sets of hyperspherical
coordinates ($\rho$,$\alpha$,$\Omega_x,\Omega_y$) are elsewhere
defined, see e.g.~\cite{fed94a,fed94b,zhu93}.  Here
$\rho~(=\sqrt{x^2+y^2})$ is the generalized radial coordinate and
$\alpha$, in the interval $[0,\pi/2]$, defines the relative size of
${\bf x}$ and ${\bf y}$ , $\Omega_x$ and $\Omega_y$ are the angles
describing the directions of ${\bf x}$ and ${\bf y}$.  One of these
sets of hyperspherical coordinates is in principle sufficient for a
complete description.  The volume element is given by $\rho^5{\rm
d}\Omega{\rm d}\rho$  where ${\rm d}\Omega=\sin^2 \alpha
\cos^2 \alpha {\rm d}\alpha {\rm d}\Omega_x {\rm d}\Omega_y$.

The total wavefunction $\Psi$ of the three-body system is written as a
sum of three components each expressed in terms of one of the three
different sets of Jacobi coordinates:
\begin{equation} \label{e4}
\Psi= \sum_{i=1}^{3}  \psi^{(i)}({\bf x}_i,{\bf y}_i).
\end{equation}
This three-component wavefunction is flexible and allows a
description of different three-body structures by means of rather few
angular momenta in each component.  These wavefunctions satisfy the
three Faddeev equations \cite{tho93}
\begin{equation} \label{e5}
(T-E)\psi^{(i)} +V_{jk} (\psi^{(i)}+\psi^{(j)}+\psi^{(k)})=0,\;
\end{equation}
where $E$ is the total energy, $T$ is the kinetic energy operator and
$\{i,j,k\}$ is a cyclic permutation of $\{1,2,3\}$. Any solution to
eq.~(\ref{e5}) is via eq.~(\ref{e4}) also a solution to the
Schr\"{o}dinger equation, but the Faddeev equations, which in practice
almost inevitably appear as integro-differential equations, are much
better suited for descriptions of delicate correlations.

Each component $\psi^{(i)}$ is now for each $\rho$ expanded in a
complete set of generalized angular functions
$\Phi_n^{(i)}(\rho,\Omega_i)$.
\begin{equation} \label{e6}
	\psi^{(i)} = \frac {1}{\rho^{5/2}}
	\sum_n f_n(\rho)  \Phi_n^{(i)}(\rho ,\Omega_i)
	\; ,
\end{equation}
where $\rho^{-5/2}$ is the phase space factor.  The angular
functions must be chosen to describe correctly the asymptotic behavior
of the wavefunction and in particular the behavior corresponding to
possible bound two-body subchannels. This is essential for low
energies where the correlations must be described very accurately.
These functions are now chosen for each $\rho$ as the eigenfunctions
of the angular part of the Faddeev equations:
\begin{equation} \label{e7}
 {\hbar^2 \over 2m}\frac{1}{\rho^2}\hat\Lambda^2 \Phi_n^{(i)} +V_{jk}
(\Phi_n^{(i)}+\Phi_n^{(j)}+\Phi_n^{(k)})\equiv {\hbar^2 \over
2m}\frac{1}{\rho^2} \lambda_n (\rho) \Phi_n^{(i)} ,\;
\end{equation}
where $\{i,j,k\}$ again is a cyclic permutation of $\{1,2,3\}$ and
$\hat\Lambda^2$ is the $\rho$-independent part of the kinetic energy
operator defined by
\begin{equation}\label{e8}
      T \equiv T_{\rho}+{\hbar^2 \over 2m}\frac{1}{\rho^2}\hat\Lambda^2, \;
      T_{\rho}=-{\hbar^2 \over 2m}\left(
\rho^{-5/2}\frac{{\partial}^2}{{\partial}\rho^2}
\rho^{5/2}-\frac{1}{\rho^2} \frac{15}{4}\right) \; .
\end{equation}
Explicitly the generalized angular momentum operator $\hat \Lambda^2$
is given by
\begin{equation}  \label{e8a}
	\hat \Lambda^2=-{1 \over \sin\alpha\cos\alpha}
      {{\partial}^2\over {\partial} \alpha^2} \sin\alpha \cos\alpha
	+{\hat l_x^2 \over {\sin^2 \alpha}}
 	+{\hat l_y^2 \over {\cos^2 \alpha}} -4
\end{equation}
in terms of the angular momentum operators $\hat l_{x}^2$ and $\hat
l_{y}^2$ related to the ${\bf x}$ and ${\bf y}$ degrees of freedom.
This procedure is particularly convenient, since all angular variables
are confined to finite parameter intervals and therefore correspond to
discrete eigenvalue spectra. The idea, here applied to the Faddeev
equations, was exploited previously to solve the Schr\"{o}dinger
equation in atomic physics \cite{mac68} and recently also in studies
of the triton \cite{das93}.

The radial expansion coefficients $f_n(\rho)$ are component
independent, since $\rho$ is independent of which Jacobi coordinates are
used. Insertion of eq.~(\ref{e6}) into eq.~(\ref{e5}) and use of
eqs.~(\ref{e7}), (\ref{e8}) and (\ref{e8a}) then lead to the following
coupled set of ``radial'' differential equations:
\begin{equation} \label{e9}
   \left(-\frac{\rm d ^2}{\rm d \rho^2}
   -{2mE\over\hbar^2}+ \frac{1}{\rho^2}\left( \lambda_n(\rho) +
  \frac{15}{4}\right) \right)f_n
   +\sum_{n'}   \left(
   -2P_{nn'}{\rm d \over\rm d \rho}
   -Q_{nn'}
   \right)f_{n'}
   =0 , \;
\end{equation}
where the functions $P$ and $Q$ are the following angular integrals:
\begin{equation}\label{e10}
   P_{nn'}(\rho)\equiv \sum_{i,j=1}^{3}
   \int d\Omega \Phi_n^{(i)\ast}(\rho,\Omega)
   {\partial\over\partial\rho}\Phi_{n'}^{(j)}(\rho,\Omega),\;
\end{equation}
\begin{equation}\label{e11}
   Q_{nn'}(\rho)\equiv \sum_{i,j=1}^{3}
   \int d\Omega \Phi_n^{(i)\ast}(\rho,\Omega)
   {\partial^2\over\partial\rho^2}\Phi_{n'}^{(j)}(\rho,\Omega).\;
\end{equation}
Often very few terms of the expansion in eq.~(\ref{e6}) is needed to
get sufficient accuracy in calculations of the lowest lying states. On
the other hand the method is not suited for highly excited states
where many high angular momenta are needed in the expansion. Systems
with many crossings of the $\lambda$-values as function of $\rho$ are
also numerically difficult to solve accurately by this method.

\section{Two-body potentials}
Loosely bound systems are mainly sensitive to the low-energy
properties of the potentials and we shall therefore use relatively
simple potentials reproducing the available low-energy scattering data
and still allowing extensive three-body calculations. The
neutron-neutron interaction in the singlet s-wave is assumed to be the
same as in ref.\cite{joh90}, where the radial shape is a gaussian. In
the extension to triplet p-waves we include spin-spin, spin-orbit and
tensor terms. The resulting potential is then
\begin{equation}\label{e12}
 V_{nn} = \left( V_c + V_{ss}{\bf s}_{n1}\cdot{\bf s}_{n2} + V_T \hat S_{12}
 + V_{so}{\bf l}_{nn}\cdot{\bf s}_{nn}\right) {\rm exp}({-(r/b_{nn})^2}),\;
\end{equation}
where ${\bf s}_{n1}$ and ${\bf s}_{n2}$ are the spins of the neutrons,
${\bf s}_{nn} = {\bf s}_{n1}+ {\bf s}_{n2}$ and $\hat S_{12}$ is the
usual tensor operator. The four strength parameters and the range of
the gaussian are then adjusted to reproduce the following four
scattering lengths $a$ and the s-wave effective range $r_e$ \cite{dum83}
\begin{eqnarray}\label{e13}
  a(^1S_0) = 18.8~{\rm fm} ,\;  r_e(^1S_0) = 2.76~{\rm fm} , \; \nonumber\\
 a(^3P_0) = 3.6~{\rm fm} ,\;  a(^3P_1) = -2.0~{\rm fm} ,\;
  a(^3P_2)  = 0.30~{\rm fm} .\;
\end{eqnarray}
The resulting values are given by
\begin{eqnarray}\label{e14}
 V_c = 2.92~{\rm MeV} , \;  V_{ss} = 45.22~{\rm MeV} , \;
     V_{T} = 26.85~{\rm MeV} , \;  V_{so} = -12.08~{\rm MeV} , \;  \nonumber\\
     b_{nn} = 1.8~{\rm fm} , \;
\end{eqnarray}
which by use in eq.~(\ref{e13}) leads to the proper low-energy
behavior both in the s-wave and the three different p-waves.

The neutron-core potential, which also is assumed to have a gaussian
radial shape, is directly related to the resonances or virtual states
in the subsystem. Their positions in energy are crucial in the
computations and it is necessary to be able to adjust s-, p- and
d-waves independently. We therefore use the parametrization
\begin{eqnarray}\label{e15}
   V_{nc}^{(s)} = V_s(1+\gamma_s {\bf s}_c \cdot {\bf s}_n )
   {\rm exp}(-{(r/b_{nc})}^2) , \;  \nonumber\\
  V_{nc}^{(l)} = (V_l~+~V_{so}^{(l)} {\bf s}_n \cdot {\bf l}_{nc})
  {\rm exp}(-{(r/b_{nc})}^2)  , \;
\end{eqnarray}
where $l=1,2$ corresponds to either p- or d-waves, $V_s, V_l,
V_{so}^{(l)}, \gamma_s$ and $b_{nc}$(=2.55~fm) are constants and ${\bf
s}_c, {\bf s}_n$ and ${\bf l}_{nc}$ are the spin of the core, spin of
the neutron and the relative neutron-core orbital angular momentum.
The spin splitting is for simplicity only introduced in the s-state,
but could as well be important for the other partial waves. Both the
spin-orbit partners of higher orbital angular momenta may be
independently adjusted.  These effective potentials must take into
account that the orbits occupied by the neutrons in the core are
excluded by the Pauli principle. This is either done by choosing
shallow potentials unable to hold bound states or by considering
higher lying three-body states.

We first choose \cite{joh90} $V_s = V_p = -7.8$~MeV, $V_{so}^{(l)} =
0$ and $\gamma_s=0$, which results in a total binding energy of
0.3~MeV corresponding to $^{11}$Li.  Maintaining this binding energy
of the three-body system and varying $V_s$ and $V_{so}$, we obtain the
curve in fig.~1 relating the virtual $s_{1/2}$-state and the
$p_{1/2}$-resonance in the neutron-core system.  When one of these
states is very low the corresponding component is dominating in the
three-body wavefunction. The straight lines between the calculated
points are inserted to guide the eye.

The spin splitting is now investigated by varying $\gamma_s$ and $V_s$
while keeping the binding energy at 0.3~MeV and $V_{so}$ at the values
corresponding to the square and the vertical line in fig.~1. The
resulting connection between the virtual s-states are shown in fig.~2
for these two cases. When the energy of one of the states increases
the other decreases due to the inherent coupling in the Faddeev
equations.  A finite p-state admixture (the squares) increases the
energies of the virtual s-states to maintain the total binding energy.
In both cases the energy of one of the spin-split virtual s-states in
the neutron-core system may occur anywhere below 0.1~MeV.

The angular momentum dependence is now investigated by using a
$d_{5/2}$-resonance instead of the $p_{1/2}$-resonance in the
neutron-core system. This might be the relevant structure for
$^{19}$B, where the $^{17}$B-core probably has an essentially fully
occupied p-shell. The total angular momentum of $^{17}$B is 3/2 which
also is the expected value for $^{19}$B. This is consistent with
$l_c=l_{nn}=s_{nn}=0$, which normally would be the state of lowest
energy in our three-body calculation.  The two-neutron separation
energy of $^{19}$B is estimated to be $500 \pm 450~$~keV , see
ref.~\cite{aud93}, and could in fact be anything below 1~MeV.
Therefore we vary the energy along lines relating the neutron-core
resonance energies by E($d_{5/2}$)=xE($s_{1/2}$), where x then is a
number between 0 and $\infty$. Varying $V_s$ and $V_d$ we obtain the
total three-body energy shown in fig.~3 as function of E($s_{1/2}$)
for several values of x. The p-waves are not included in the allowed
configuration space. This is equivalent to a choice of p-strength
$V_p$, which places the p-states at high positive energies. The spin
orbit strength $V_{so}^{(d)}=-10$~MeV and the spin splitting parameter
$\gamma_s=0$ are kept constant in the calculation. The starting point
is $V_s = -7.87$~MeV or $V_s = -7.40$~MeV corresponding to a virtual
s-state at 50~keV or 100~keV, respectively.  The total three-body
energy decreases almost linearly with increasing resonance energies
for all x-values and remains in all the cases considered within the
limits of the estimate.

The total binding energies for the black triangles in fig.~3 are
around 0.45~MeV and 0.75~MeV.  The effect of spin splitting in the
s-state is now investigated by increasing $\gamma_s$ from zero while
keeping the other parameters at values corresponding to these points.
The strengths of the two potentials for a total spin of $s_c\pm 1/2$
are proportional to the related expectation values $\langle
1+\gamma_s{\bf s}_c \cdot {\bf s}_n \rangle$, which respectively is
$1+\gamma_s s_c/2$ or $1-\gamma_s (s_c+1)/2$, see eq.~(\ref{e15}). The
effective radial three-body potential is for small $\rho$ a linear
combination of these two two-body potentials with their respective
statistical weights and it converges for large $\rho$ towards the
hyperspherical spectrum, i.e.~values independent of the potential.
This radial potential is consequently almost independent of $\gamma_s$
and the resulting three-body energy is then also almost independent of
$\gamma_s$. The connection between the energies of the two virtual
s-states corresponding to the spins $s_{nc}=1,2$ of the neutron-core
system is shown in fig.~4 as function of $\gamma_s$ for interaction
parameters corresponding to the black triangles in fig.~3. The
qualitative behavior is similar to that of fig.~2, since the total
energy remains approximately unchanged.

\section{Angular eigenvalue equations for s-states}
As seen from eq.~(\ref{e9}) the angular eigenfunction $\lambda (\rho)$
is essentially an effective potential determined by eq.~(\ref{e7}).
The generalized angular functions $\Phi_n^{(i)}(\rho,\Omega_i)$ is a
sum of components with different angular momenta of the subsystems
coupled to the total angular momentum. The radial equations at large
$\rho$ do not couple neither these components nor those belonging to
different Jacobi coordinates with the important exception of the
states with zero two-body orbital angular momentum. These s-states are
often essential components in low lying states of interest and we
furthermore only introduced the spin splitting in s-states. We shall
therefore first concentrate on s-states where the angular dependence
in eq.~(\ref{e6}) is reduced to contain only $\alpha$.

The spin dependence of the i'th component of the total wave function
in eq.~(\ref{e4}) is denoted $\chi_{s}^{(i)}$, where ${\bf s}$ is the
intermediate spin obtained by coupling of the spins (${\bf s}_j$ and
${\bf s}_k$) of particles $j$ and $k$. The spin ${\bf s}$ is afterwards
coupled to ${\bf s}_i$ to give the spin of the core ${\bf s}_c$ which
in our case equals the total spin of the system.  For each channel the
angular part can then be written as
\begin{eqnarray}\label{e16a}
\Phi^{(1)}(\rho,\alpha_1) & = &
     \frac{\phi_{0}^{(1)}(\rho,\alpha_1)}{\sin{2 \alpha_1}}
 \chi_{0}^{(1)}   , \;
\end{eqnarray}
\begin{eqnarray}\label{e16b}
\Phi^{(2)}(\rho,\alpha_2) & = &  \frac{\phi_{s_c-1/2}^{(2)}(\rho,\alpha_2)}
 {\sin{2 \alpha_2}} \chi_{s_c-1/2}^{(2)}
 +\frac{\phi_{s_c+1/2}^{(2)}(\rho,\alpha_2)}{\sin{2 \alpha_2}}
 \chi_{s_c+1/2}^{(2)}  , \;
\end{eqnarray}
\begin{eqnarray}\label{e16c}
\Phi^{(3)}(\rho,\alpha_3) & = &
 \frac{\phi_{s_c-1/2}^{(3)}(\rho,\alpha_3)}{\sin{2 \alpha_3}}
 \chi_{s_c-1/2}^{(3)}
 +\frac{\phi_{s_c+1/2}^{(3)}(\rho,\alpha_3)}{\sin{2\alpha_3}}
 \chi_{s_c+1/2}^{(3)}  , \;
\end{eqnarray}
where the channel i=1 is defined by {\bf x} between the two neutrons
and where $\phi_{s}^{(i)}$ is the spatial part of the wavefunction related
to $\chi_{s}^{(i)}$. Since the total wave function must be
antisymmetric under interchanging of the two neutrons we have
$\phi_{s_c-1/2}^{(3)} = \phi_{s_c-1/2}^{(2)}$ and
$\phi_{s_c+1/2}^{(3)} = - \phi_{s_c+1/2}^{(2)}$.

Substituting eqs.(\ref{e16a})--(\ref{e16c}) into eq.~(\ref{e7}) with
$i=1,2$ and subsequent multiplication from the left by
$\chi_{0}^{(1)}$, $\chi_{s_c-1/2}^{(2)}$ and $\chi_{s_c+1/2}^{(2)}$,
we obtain the following set of equations for the functions
$\phi_{0}^{(1)}(\rho,\alpha_1)$, $\phi_{s_c-1/2}^{(2)}(\rho,\alpha_2)$ and
$\phi_{s_c+1/2}^{(2)}(\rho,\alpha_2)$:
\begin{small}
\begin{eqnarray}\label{e17a}
&&\left(
   -\frac{1}{\sin{2\alpha_1}}\frac{\partial^2}{\partial \alpha_1^2}
\sin{2\alpha_1} - \tilde{\lambda}(\rho)
\right) \frac{\phi_{0}^{(1)}(\rho,\alpha_1)}{\sin{2\alpha_1}} +
\rho^2 v_{nn}(\rho \sin{\alpha_1})
\left[ \frac{\phi_{0}^{(1)}(\rho,\alpha_1)}{\sin{2\alpha_1}} +
 \right.  \\
&& \left.
 C_1 \frac{\phi_{s_c-1/2}^{(2)}(\rho,\alpha_2)}{\sin{2\alpha_2}}
+C_2 \frac{\phi_{s_c+1/2}^{(2)}(\rho,\alpha_2)}{\sin{2\alpha_2}} +
C_1
\frac{\phi_{s_c-1/2}^{(2)}(\rho,\alpha_3)}{\sin{2\alpha_3}}
+C_2 \frac{\phi_{s_c+1/2}^{(2)}(\rho,\alpha_3)}{\sin{2\alpha_3}} \right]= 0  ,
\; \nonumber
\end{eqnarray}

\begin{eqnarray}\label{e17b}
&&\left(
   -\frac{1}{\sin{2\alpha_2}}\frac{\partial^2}{\partial \alpha_2^2}
\sin{2\alpha_2}- \tilde{\lambda}(\rho)
\right) \frac{\phi_{s_c-1/2}^{(2)}(\rho,\alpha_2)}{\sin{2\alpha_2}} +
\rho^2 \ \langle \chi_{s_c-1/2}^{(2)} |v_{nc}(\rho \sin{\alpha_2})
 | \chi_{s_c-1/2}^{(2)} \rangle \nonumber \\ &&
\left[  C_1
\frac{\phi_{0}^{(1)}(\rho,\alpha_1)}{\sin{2\alpha_1}}+
\frac{\phi_{s_c-1/2}^{(2)}(\rho, \alpha_2)}{\sin{2\alpha_2}} +
 C_3 \frac{\phi_{s_c-1/2}^{(2)}(\rho,\alpha_3)}{\sin{2\alpha_3}}
-C_4  \frac{\phi_{s_c+1/2}^{(2)}(\rho,\alpha_3)}{\sin{2\alpha_3}} \right]=0  ,
\;
\end{eqnarray}

\begin{eqnarray}\label{e17c}
&&\left(
   -\frac{1}{\sin{2\alpha_2}}\frac{\partial^2}{\partial \alpha_2^2}
\sin{2\alpha_2}- \tilde{\lambda}(\rho)
\right) \frac{\phi_{s_c+1/2}^{(2)}(\rho,\alpha_2)}{\sin{2\alpha_2}} +
\rho^2 \ \langle \chi_{s_c+1/2}^{(2)} |v_{nc}(\rho \sin{\alpha_2})
 | \chi_{s_c+1/2}^{(2)} \rangle \nonumber \\ &&
\left[C_2
\frac{\phi_{0}^{(1)}(\rho,\alpha_1)}{\sin{2\alpha_1}}+
\frac{\phi_{s_c+1/2}^{(2)}(\rho, \alpha_2)}{\sin{2\alpha_2}}
 -C_4 \frac{\phi_{s_c-1/2}^{(2)}(\rho,\alpha_3)}{\sin{2\alpha_3}}
-C_3  \frac{\phi_{s_c+1/2}^{(2)}(\rho,\alpha_3)}{\sin{2\alpha_3}} \right]=0  ,
\;
\end{eqnarray}
\end{small}
where $v_{ij}(x)= V_{ij}(x/a_{ij}) 2m / \hbar^2$,
$a_{jk} = \left[ m_j m_k/(m(m_j+m_k)) \right]^{1/2}$,
$\tilde{\lambda}(\rho) =
\lambda(\rho)+4$ and where we used that the neutron-neutron interaction
($v_{nn}=v_{23}$) is spin independent for s-waves and that the
neutron-core interaction ($v_{nc}=v_{12}=v_{13}$) is diagonal in spin
space.  The coefficients $C_i$, i=1,2,3,4 depend on the spin of the
core, and take the values: 
\begin{equation}\label{e18}
\begin{array}{l}
C_1= \langle \chi_{0}^{(1)} | \chi_{s_c-1/2}^{(2)} \rangle =
      \langle \chi_{0}^{(1)} | \chi_{s_c-1/2}^{(3)} \rangle =
 -\sqrt{s_c/(2s_c+1)} \\
C_2=  \langle \chi_{0}^{(1)} | \chi_{s_c+1/2}^{(2)} \rangle =
    -  \langle \chi_{0}^{(1)}| \chi_{s_c+1/2}^{(3)} \rangle =
 \sqrt{(s_c+1)/(2s_c+1)} \\
C_3=  \langle \chi_{s_c-1/2}^{(2)} | \chi_{s_c-1/2}^{(3)} \rangle =
      \langle \chi_{s_c+1/2}^{(2)} | \chi_{s_c+1/2}^{(3)}\rangle =
 - 1/(2s_c+1)
\\
C_4= \langle \chi_{s_c-1/2}^{(2)} | \chi_{s_c+1/2}^{(3)} \rangle =
    - \langle \chi_{s_c+1/2}^{(2)} | \chi_{s_c-1/2}^{(3)} \rangle =
 \sqrt{4s_c(s_c+1)}/(2s_c+1)   . \;
\end{array}
\end{equation}
Eqs.(\ref{e17b}) and (\ref{e17c}) are identical to the
ones obtained by substituting eqs.(\ref{e16a})--(\ref{e16c}) into
eq.(\ref{e7}) with i=3 and the subsequent multiplication from the left
by $\chi_{s_c-1/2}^{(3)}$ and $\chi_{s_c+1/2}^{(3)}$.

The three spatial wavefunctions $\phi_0^{(1)}$ and $\phi_{s_c \pm
1/2}^{(2)}$ are now for each $\rho$ determined by
eqs.(\ref{e17a})--(\ref{e17c}) and the following boundary conditions
\begin{equation}\label{e19}
\phi_0^{(1)}(\rho ,\alpha =0)=\phi_0^{(1)}(\rho ,\alpha =\pi/2)=
\phi_{s_c \pm 1/2}^{(2)}(\rho ,\alpha =0) =
\phi_{s_c \pm 1/2}^{(2)}(\rho ,\alpha =\pi/2) = 0  . \;
\end{equation}
For $\rho=0$ the eqs.(\ref{e17a})--(\ref{e17c}) clearly decouple and
the solutions are
\begin{equation}\label{e20}
   \phi_{s_c \pm 1/2}^{(2)}(\rho =0,\alpha)=\phi_{0}^{(1)}(\rho
=0,\alpha) \propto \sin{(2n \alpha)}, \hspace{1cm} n=1,2,3,\cdots \;
\end{equation}
and the eigenvalues are given by
\begin{equation} \label{e21}
 \lambda_0 \equiv \lambda(\rho=0)=\tilde{\lambda}(\rho=0)-4=K(K+4),
\hspace{1cm} K=2n-2 , \;
\end{equation}
which is the well known hyperspherical spectrum.

The behavior of $\lambda$ for small $\rho$-values can now be obtained
in first order perturbation theory.  The easiest is to add the three
equations in eqs.(\ref{e17a})--(\ref{e17c}) and thereby obtain the
equivalent Schr\"{o}dinger equation, which in turn then gives the
following first order correction to $\lambda_0$
\begin{small}
\begin{eqnarray}\label{e21a}
\lambda(\rho)&=& \lambda_0 + \rho^2 \left[ v_{nn}(\rho=0) +
 \frac{2s_c}{2s_c+1} \ \langle \chi_{s_c-1/2}^{(2)} |v_{nc}(\rho=0) |
 \chi_{s_c-1/2}^{(2)}   \rangle  \right. \nonumber \\
 &&+ \left. \frac{2s_c+2}{2s_c+1} \ \langle \chi_{s_c+1/2}^{(2)}
|v_{nc}(\rho=0)
  |    \chi_{s_c+1/2}^{(2)}  \rangle
                                    \right] , \;
\end{eqnarray}
\end{small}
where we used the spin independence of the neutron-neutron
interaction. Eq.(\ref{e21a}) reduces for vanishing core-spin $s_c=0$
and spin independent interactions to the usual small $\rho$
expression, where only the sum of the interactions enters, see
ref.\cite{fed94b}.

\section{The Efimov effect}
The large distance behavior of the $\lambda$-values reflect
characteristic properties of the three-body system and its subsystems.
In particular the necessary and sufficient condition for the
occurrence of the Efimov states is simply that
$\lambda_{\infty}(\equiv \lambda(\rho \rightarrow \infty))$ is smaller
than $-4$ or equivalently $\tilde{\lambda} < 0$. Therefore eigenvalues
and eigenfunctions for large $\rho$ must be obtained from
eqs.(\ref{e17a})--(\ref{e17c}) and we shall follow the procedure in
ref.\cite{fed93b}.

The wavefunctions depending on $\alpha _k$ are first expressed in
terms of the i'th set of hyperspherical coordinates and the equation
are subsequently integrated over the four angular variables ${\bf
\Omega }_{x_k}$ and ${\bf \Omega }_{y_k}$. This amounts for s-waves to
the substitution formally expressed by the operator $R_{ik}$ defined by
\begin{equation}\label{e30}
R_{ik}\psi =
\frac{1}{\sin(2\varphi _{ik} )}\frac{1}{\sin(2\alpha _i)}
\int_{|\varphi _{ik}-\alpha _i|}^{\pi/2 - |\pi/2-\varphi _{ik} -\alpha _i|}
\sin(2\alpha _k) \psi(\rho,\alpha _k)d\alpha _k \; ,
\end{equation}
where the transformation angle $\varphi_{ik}$ is given by the masses as
\begin{equation}\label{e31}
\varphi_{ik}=\arctan\left((-1)^p\sqrt{m_j(m_1+m_2+m_3)\over m_i m_k}\right)
   \; ,
\end{equation}
where $(-1)^p$ is the parity of the permutation $\{i,k,j\}$.  The
short ranges of the potentials confine $\alpha_k$ to be smaller than the
range $r_0$ of the potential divided by $\rho$. (The precise
definition of $r_0$ is given by $\rho^2 v_{nc}(r_0) = \tilde\lambda$).
The effect of the R-operator is then for large $\rho$ approximately
given as
\begin{equation}\label{e32}
R_{ik}\psi \approx \frac{2\alpha}{\sin(2\varphi _{ik} )}
 \psi(\rho,\varphi _{ik})  \; .
\end{equation}

The wavefunction $\phi_{0}^{(1)}$ vanishes faster than $\phi_{s_c \pm
1/2}^{(2)}$ provided the scattering lengths $a_{\pm}$ related to the
two relative spin states of the neutron-core system both are larger
than that of the neutron-neutron system. We can therefore omit
eq.(\ref{e17a}) and neglect $\phi_{0}^{(1)}$ in eqs.(\ref{e17b}) and
(\ref{e17c}) and thereby for $\rho \gg r_0$ to first order in
$\alpha_0 \equiv r_0/\rho$ obtain the simplified eigenvalue equations
\begin{small}
\begin{eqnarray}\label{e32b}
 \left( -\frac{\partial^2}{\partial \alpha_2^2}
- \tilde{\lambda}(\rho) \right)
\phi_{s_c-1/2}^{(2)}(\rho, \alpha_2)  +
\rho^2 \ \langle \chi_{s_c-1/2}^{(2)} |v_{nc}(\rho \sin{\alpha_2}) |
                  \chi_{s_c-1/2}^{(2)} \rangle  \nonumber  \\
\left[\phi_{s_c-1/2}^{(2)}(\rho, \alpha_2) + \frac{2 \alpha_2}{\sin{2 \varphi}}
(C_3 \phi_{s_c-1/2}^{(2)}(\rho, \varphi) - C_4 \phi_{s_c+1/2}^{(2)}(\rho,
\varphi) )
    \right] =0    \; ,
\end{eqnarray}

\begin{eqnarray}\label{e33}
 \left( -\frac{\partial^2}{\partial \alpha_2^2}
- \tilde{\lambda}(\rho) \right)
\phi_{s_c+1/2}^{(2)}(\rho, \alpha_2) +
\rho^2 \ \langle \chi_{s_c+1/2}^{(2)} |v_{nc}(\rho \sin{\alpha_2}) |
                 \chi_{s_c+1/2}^{(2)}  \rangle  \nonumber \\
\left[\phi_{s_c+1/2}^{(2)}(\rho, \alpha_2) + \frac{2 \alpha_2}{\sin{2 \varphi}}
(-C_4 \phi_{s_c-1/2}^{(2)}(\rho, \varphi) -
C_3 \phi_{s_c+1/2}^{(2)}(\rho, \varphi) )
   \right] = 0    \; ,
\end{eqnarray}
\end{small}
where $\varphi={\rm arctan}\sqrt{A_c(A_c+2)}$ is given in terms of the
nucleon number $A_c$ of the core.

We now proceed by dividing the $\alpha _2$-space into two regions I
and II.  In region I ($\alpha _2> \alpha _0 = r_0/\rho$), where the
potential is negligibly small, the solutions to the previous equations
are:
\begin{equation}\label{e34}
 \phi_{s_c \pm 1/2}^{(2,I)}(\rho, \alpha_2)
 \propto \sin{\left((\alpha_2-\frac{\pi}{2})
\sqrt{\tilde{\lambda}} \right)}   \; ,
\end{equation}
where we have imposed that the solution has to vanish at the boundary
$\alpha_2=\pi/2$.

In region II ($\alpha _2 < \alpha _0$), where $\tilde\lambda/\rho^2$ is
negligibly small, we have two inhomogeneous differential equations
with the solutions:
\begin{small}
\begin{equation}\label{e35}
\phi_{s_c- 1/2}^{(2,II)}(\rho, \alpha_2)=u_-(\rho \alpha_2)-
 \frac{2 \alpha_2}{\sin{2\varphi}}
\left( C_3 \phi_{s_c- 1/2}^{(2,I)}(\rho,\varphi) - C_4
 \phi_{s_c+ 1/2}^{(2,I)}(\rho, \varphi) \right)   \; ,
\end{equation}
\begin{equation}\label{e36}
\phi_{s_c+ 1/2}^{(2,II)}(\rho, \alpha_2)=u_+(\rho \alpha_2)- \frac{2
\alpha_2}{\sin{2\varphi}}
\left( -C_4 \phi_{s_c- 1/2}^{(2,I)}(\rho,\varphi) - C_3
 \phi_{s_c+ 1/2}^{(2,I)}(\rho,\varphi) \right) \; ,
\end{equation}
\end{small}
where the inhomogeneous solutions, proportional to $\alpha_2$, easily
are found due to the linear dependence of $\alpha_2$. The homogeneous
solutions $u_{\pm}(\rho
\alpha_2)$ are the wave functions describing the state of zero energy
in their respective potentials $\langle \chi_{s_c\pm 1/2}^{(2)}
|v_{nc}(\rho \alpha_2) | \chi_{s_c \pm 1/2}^{(2)} \rangle$.  Outside
the range of the potential ($\rho \alpha_2>r_0$) these wavefunctions
take the form $u_{\pm}(\rho \alpha_2) \propto \rho \alpha_2 +
a_{\pm}$, where $a_{\pm}$ are the scattering lengths of the two
neutron-core potentials corresponding to the two different spin
couplings.

The eigenvalue equation for $\tilde{\lambda}$ now arises by matching
of the derivative of the logarithm of the solutions at the division
point $\alpha_2=\alpha_0=r_0/\rho$. Keeping only the lowest order in
$r_0/\rho$, and assuming that both scattering lengths are much larger
than $r_0$ ($a_{\pm} \gg r_0$), we finally obtain
\begin{small}
\begin{eqnarray}\label{e37}
&&
\tilde{\lambda} \cos^2{(\frac{\pi}{2} \sqrt{\tilde{\lambda}})} +
\frac{\rho^2}{a_- a_+} \sin^2{(\frac{\pi}{2} \sqrt{\tilde{\lambda}})} +
\frac{1}{2} (\frac{\rho}{a_-}+\frac{\rho}{a_+})
\sqrt{\tilde{\lambda}} \sin{(\pi \sqrt{\tilde{\lambda}})}  \\ &&
+ \frac{1}{2s_c+1} (\frac{\rho}{a_-}-\frac{\rho}{a_+}) \frac{2}{\sin{2
\varphi}}
\sin{(\frac{\pi}{2} \sqrt{\tilde{\lambda}})}
\sin{((\varphi-\frac{\pi}{2}) \sqrt{\tilde{\lambda}})}
-\frac{4}{\sin^2{2\varphi}} \sin^2{((\varphi-\frac{\pi}{2})
\sqrt{\tilde{\lambda}})} = 0 \; .
  \nonumber
\end{eqnarray}
\end{small}
The dependence on the core-spin $s_c$ appears only through the
scattering lengths $a_{\pm}$ and the parameter $C_3=-1/(2s_c+1)$.

When one or both of the scattering lengths are finite and negative,
which means that at least one bound state exists in the neutron-core
subsystem, the lowest $\tilde{\lambda}$-eigenvalue diverges as
$-\rho^2/a^2$, where $a$ is the scattering length for the channel of
strongest binding.  When both $a_{\pm}$ are positive or infinitely
large, we can distinguish between three different cases. In the first
case both $a_+$ and $a_-$ are infinitely large. Then by keeping only the
leading order, eq.(\ref{e37}) reduces to
\begin{equation}\label{e38}
\tilde{\lambda} \cos^2{(\frac{\pi}{2} \sqrt{\tilde{\lambda}}) } -
 \frac{4}{\sin^2{2 \varphi}}
\sin^2{((\varphi-\frac{\pi}{2})\sqrt{\tilde{\lambda}})} = 0 \; ,
\end{equation}
which for $^{11}$Li, where $\varphi=\arctan{\sqrt{11\cdot 9}}$, has the lowest
solution at $\tilde{\lambda}=-0.00545$. This value gives rise to an
effective attractive potential in the radial equation eq.(\ref{e9})
given by $-0.25545/\rho^2$, which results in an infinite number of
bounds states.  This number is independent of the core-spin $s_c$, and
the only dependence on the nucleus is contained in the angle $\varphi$,
see eq.(\ref{e31}).

In the opposite limit, where $\rho$ is large compared to both
scattering lengths, i.e.~$a_{\pm} \ll \rho$, the leading order
approximation of eq.(\ref{e37}) is simply
\begin{equation}\label{e39}
\sin^2{(\frac{\pi}{2} \sqrt{\tilde{\lambda}})} = 0   \;
\end{equation}
and the hyperspherical spectrum is recovered, i.e.~$\tilde{\lambda}-4
= K(K+4), K=0,2,4..$.

The last case is characterized by one small and one large scattering
length compared to $\rho$, i.e.~$0 < a_- \ll \rho \ll a_+$, or $0 <
a_+ \ll
\rho \ll a_-$. Now eq.(\ref{e37}) reduces to leading order to the form
\begin{equation}\label{e40}
\sin{(\frac{\pi}{2} \sqrt{\tilde{\lambda}})} \left[
  \sqrt{\tilde{\lambda}} \cos{(\frac{\pi}{2}  \sqrt{\tilde{\lambda}})}
\pm \frac{1}{2s_c+1} \frac{2}{\sin{2\varphi}}
  \sin{((\varphi-\frac{\pi}{2}) \sqrt{\tilde{\lambda}})}   \right] = 0 \; ,
\end{equation}
where the $\pm$ signs appear for $a_{\mp} \ll \rho \ll a_{\pm}$.  This
equation is independent of $a_{\pm}$ and it has never any negative
solutions.  The lowest positive solutions for $^{11}$Li are
$\tilde{\lambda}=0.702627$ and $\tilde{\lambda} = 1.34930$,
respectively when $a_- \ll \rho \ll a_+$ and $a_+ \ll \rho \ll a_-$.
Since $\tilde\lambda$ always is positive this means that the barrier
in the radial equation ($(\tilde\lambda - 1/4)/\rho^2$) never can give
rise to the Efimov effect. This reflects the fact that infinitely many
bound states only occur when at least two of the subsystems have
infinitely large scattering lengths. If this extreme effect should be
present both $a_{\pm}$ must be infinitely large.

The behavior of the $\tilde\lambda$ function is summarized in fig.~5,
where the lowest $\tilde\lambda$ for $^{11}$Li is shown. The
scattering lengths have been chosen to be $10^2$~fm and $10^7$~fm, and
the dashed (solid) line corresponds to $a_- < a_+$ ($a_+ < a_-$). When
both $a_-$ and $a_+$ are much larger than $\rho$, both curves converge
to the same value of $\tilde\lambda$ = $-0.00545$. In the opposite
limit where $\rho$ is larger than both scattering lengths, the curves
both approach the value four corresponding to the lowest level of the
hyperspherical spectrum.  In the intermediate region, where $\rho$ is
much larger than one of the scattering lengths and much smaller than
the other, $\tilde\lambda$ has a flat region.  The constant value
depends in this case only on which of the scattering lengths is larger
than the other, but not on the specific values of $a_-$ and $a_+$.
These $\tilde\lambda$ values are shown as dashed lines in the figure.

It is important to emphasize that the conditions for the appearance of
the Efimov states become more restrictive for a finite spin of the
core. In principle, the effect appears when at least two of the binary
subsystems have extremely large scattering lengths. We have seen above
that both the possible couplings of the neutron-core spin states for
halo nuclei must correspond to extremely large scattering lengths.  If
only one of the scattering lengths is very large the $\lambda$
function is not low enough to permit many bound states, i.e.
occurrence of the Efimov effect. However, these conclusions do not
affect halo nuclei with zero core spin. It should still be remembered
that the most promising place in nuclei to look for the Efimov effect
would be in a system, where one neutron is added to a pronounced
one-neutron halo nucleus with zero core spin \cite{fed94b}.

\section{Spatial structure}
The angular eigenvalues $\lambda$ are the effective radial potentials.
The spectrum is therefore decisive for the spatial structure of the
system. In fig.~6 is shown an example for interaction parameters
corresponding to the square in fig.~1, i.e.~realistic values for
$^{11}$Li.  The beginning of the hyperspherical spectrum ($K(K+4),
K=0,2,4...$) is seen at both $\rho = 0$ and at $\rho = \infty$.  The
lowest level decreases from zero at $\rho = 0$, goes through a minimum
and returns back to zero at infinity. The return to zero at large
distances means that none of the two-body interactions is able to hold
bound states.  The negative pocket at smaller distances signifies an
attraction in the total three-body system. The attraction is in this
case able to hold the system in a bound state although no binary
subsystem is bound.

The angular structure corresponding to the eigenvalue spectrum is
conveniently discussed in the Jacobi coordinate set where ${\bf x}$ is
the vector between the two neutrons. The components of the
wavefunction for $\rho = 0$ and $\rho = \infty$ corresponds to
$(l_{nn},l_c,L,s_{nn},s_c,S)$ = $(0,0,0,0,3/2,3/2)$ for the lowest
level. The five fold degeneracy of the $K=2$ levels corresponds to the
angular components $(l_{nn},l_c,L,s_{nn},s_c,S)$ = $(0,0,0,0,3/2,3/2)$
with one node, $(1,1,0,1,3/2,3/2)$ and the three levels
$(1,1,1,1,3/2,S)$ with $S=1/2,3/2,5/2$.

The usual measure of the size of the system is the mean square radius,
which for $A_c$ nucleons in the core is given by
\begin{equation}\label{e25}
   R_{RMS}^2 = \frac{A_c}{A_c+2}R_{RMS}^2({\rm core}) +
  \frac{1}{A_c+2}<\rho^2> , \;
\end{equation}
where $R_{RMS}^2$ and $R_{RMS}^2({\rm core})$ are mean square radii of
the respective systems and the mean square of the hyperradius is given
by $<\rho^2>= A_c<r_c^2>+2<r_n^2>$, where $r_c$ and $r_n$ are core-
and neutron-coordinates measured from the center of mass of the total
system.

In fig.~7 we show the mean square radii along the curves in figs.~2
and 4 as functions of energy of one of the virtual s-states. Clearly
this measure of the extension of the system is essentially independent
of the spin splitting in all cases. This is due to the fact that the
spin splitting gives rise to very small changes of the angular
eigenvalue spectrum. The lowest $\lambda$ functions are
indistinguishable at large distances and only small differences occur
at the smaller distances, where the wavefunction has small
contributions.

However, the size depends both on the binding energy and the amount of
higher angular momentum states in the neutron-core motion. For a given
binding energy, the energy of the virtual s-state decreases with
increasing s-state content in the neutron-core subsystem, and then the
spatial extension of the system increases due to the smaller
centrifugal barrier.  The essential parameter is the amount of s-wave
in the relative neutron-core motion.  The variation of the root mean
square radius of $^{11}$Li and $^{19}$B as functions of the s-wave
probability along is shown in fig.~8.  The white circles represent the
variation of the radius for $^{11}$Li along the curve in fig.~1.  The
black circles represent the same variation for $^{19}$B, but along the
horizontal dashed line in fig.~3, which corresponds to a constant
energy of around 0.65~MeV.  The resulting almost linear dependence for
$^{11}$Li is varying within limits consistent with both computations
\cite{zhu93} and the recent interpretation ($R_{RMS}(^{11}{\rm Li}) =
3.1 \pm 0.3$~fm) of the measured reaction cross sections \cite{tan92}.

In the ground state the ratio between the weights of the spin split
s-waves are predefined by the symmetry constraint in such a way that
the parametrization in eq.(\ref{e15}) makes the $\lambda$ values
nearly independent of $\gamma_s$.  The resulting spatial three-body
structure is therefore independent of the spin splitting provided the
total energy and thereby the total s-state occupation probability
remains unchanged. Any observable for the three-body system depending
solely on the spatial structure is therefore unable to determine the
spin splitting of the subsystem. Direct measurements of the
neutron-core virtual states would of course in principle be able to
give the information. However, angular momentum determination of these
unstable nuclear states are needed.

A recent suggestion to determine the s-state admixture in the
three-body ground state of $^{11}$Li is highly relevant in this
connection \cite{zin94}. The idea is that in a nuclear break-up
reaction the final-state interaction between the core and the neutron
is decisive for the observed neutron momentum distribution. Low lying
virtual s-states or equivalently strongly attractive s-state
potentials will via the final state interaction lead to a narrow
momentum distribution \cite{bar93}. A higher lying s-state or a
p-state with repulsive centrifugal barrier can only weakly influence
the outgoing neutron. The necessary analyses require first of all the
probability of finding the remaining neutron-core system in the low
lying s-state.  For a non-negligible spin splitting this probability
equals the probability of finding two occupied s-states multiplied by
the probability of survival (roughly the statistical weights
independent of the spin splitting) of the low lying state. Therefore
information about the spin splitting may be obtained from these
reactions.

\section{Magnetic dipole moment}
Structural changes due to the spin splitting are most likely to show
up in spin dependent observables. The natural first choice to study is
the magnetic dipole moment, which in our three-body model is given by
\begin{equation}\label{e22}
 \frac{\mu}{\mu_N} =\langle JM=J|(g_n{\bf s}_{nn}+g_c{\bf s}_c+
   Z_c\frac{2}{A_c+2}{\bf l}_c)_z|JM=J\rangle ,\;
\end{equation}
where $\mu_N$ is the nuclear magneton, $Z_c$ is the charge of the
core, ${\bf l}_c$ is the orbital angular momentum of the core relative
to the center of mass of the neutrons, J and M are the total angular
momentum and its projection. The gyromagnetic factors are taken as
$g_n=-3.8263$ for the neutron, $g_c=3.4391/s_c$ for $^{11}$Li and the
Schmidt value $g_c=3.7928/s_c$ for $^{19}$B.  The non-vanishing matrix
elements of the operator in eq.~(\ref{e22}) are then
\begin{eqnarray}\label{e23}
&\langle(l_{nn}l_c)L(s_{nn}s_c)S;JJ|~({\bf s}_{nn})_z~
	|(l_{nn}l_c)L(s_{nn}s_c)S';JJ\rangle=
\nonumber\\
& (-1)^{J+L+S+S'+s_{nn}+s_c}
\sqrt{J(2J+1)(2S+1)(2S'+1)s_{nn}(s_{nn}+1)(2s_{nn}+1)/(J+1)}
\nonumber\\
&\times
\left\{\begin{array}{ccc} S&1&S'\\ J&L&J \end{array}\right\}
\left\{\begin{array}{ccc} s_{nn}&1&s_{nn}\\ S'&s_c&S \end{array}\right\},\;
\end{eqnarray}
where L is the total orbital angular momentum, $l_{nn}$ is the orbital
angular momentum of the two neutrons, S (S') is the total spin.  The
expression for the other two terms of the operator in eq.~(\ref{e22})
can be obtained respectively by exchange of ${\bf s}_c$ and ${\bf
s}_{nn}$, and by additional exchanges of all spins and orbital angular
momenta on the right hand side of the equation.

The results of these formulae are given in table~1 for both our
examples of $^{11}$Li and $^{19}$B. The major contributions consistent
with parity and angular momentum coupling rules are listed in the
neutron-neutron Jacobi set of coordinates.  All diagonal p-wave matrix
elements except one are smaller than the s-wave matrix element.  Since
all these components are coupled in the three-body system one could
expect that p-wave admixture will reduce the magnetic moment compared
to the core-value in contradiction with measurements for $^{11}$Li
\cite{arn87}.  However, the diagonal elements alone are very
misleading due to the strong off diagonal matrix elements given at the
bottom of the table.

The result of direct numerical calculations for both $^{11}$Li and
$^{19}$B show that the core-value of the magnetic moment always is
recovered and it is therefore insensitive both to the spin splitting
and to the amount of p- or d-wave admixture.  The reason is that the
basic neutron-core components in the present three-body model for
$^{11}$Li are the $({\rm s}_{1/2})^2$ and $({\rm p}_{1/2})^2$
two-particle states. Both are fully occupied j-shells of total angular
momentum zero. Their contributions to the magnetic moment therefore
must vanish and the off diagonal elements are essential in this
connection.  We are therefore left with the core-value.  Our model
also allows other configurations, which however turn out to be
essentially unoccupied in the final wavefunction. For $^{19}$B, where
the p-states are substituted by d-states, we also recover the
core-value for the magnetic moment. The reason is that the new
configuration $({\rm d}_{5/2})^2$ almost exclusively couples to the
total spin zero and therefore also in this case only contributes
insignificantly. Other components are apparently energetically too
expensive.

The accuracy of the strict three-body model is relying on the
assumption of an inert core. Thus core degrees of freedom must be
included when improvements beyond the three-body model are requested.
The most obvious modification for $^{11}$Li is to allow an admixture of the
only particle stable excited state, an angular momentum 1/2-state
\cite{ajz88}, into the 3/2-~ground state of $^9$Li. Since all matrix
elements of the magnetic dipole operator in eq.~(\ref{e22}) vanish
between states of different core-spin, only the contributions in
table~2 are able to modify our previous results.  Although we assumed
an unchanged gyromagnetic factor for the core, it is rather obvious
that any $s_c=1/2$ admixture would further reduce the magnetic moment.
A value of $s_c$ larger than 3/2 is needed, but not available in the
discrete spectrum of $^9$Li.  The effect of higher angular momentum
neutron-core states and a j-shell only partially occupied by the halo
neutrons does not alter the conclusions above as seen from our example
of $^{19}$B.

\section{Excited states of electric dipole character}
The spectrum of excited states in halo nuclei is almost by definition
limited to at most a few discrete states, but the continuum structure
including resonances might also be interesting. The $1^-$ type of
excitations, continuous or discrete, are for such nuclei expected to
differ significantly from corresponding states in normal nuclei.  The
only excited state in $^{11}$Be has the largest measured B(E1)-value to
the ground state \cite{mil83}. In general the dipole strength function
is peaked at low excitation energies and it plays a dominant role in
the analysis of Coulomb dissociation cross sections
\cite{han87,suz90,dan94}. The peak has even been suggested to be a new
type of low lying excitation mode for halo nuclei \cite{fay91,esb92}.
In any case it can be expected that these $1^-$ excitations are the
lowest lying excitations in halo nuclei.

When the total angular momentum as well as the spin of the core both
are zero the major components in the wavefunction of the excited
$1^--$state are most conveniently described in the neutron-neutron
coordinate system. The angular momentum 1 and the negative parity can
be made by $(l_{nn},l_c,L,s_{nn},s_c) = (0,1,1,0,0)$ and $(1,0,1,1,0)$
if we for simplicity exclude d- and higher partial waves. The finite
core spin now dictates the additional coupling of $s_c$ to a total
angular momentum of $J = s_c, s_c \pm 1$. The parity is still the
opposite of the ground state.

For a core spin of $3/2$ corresponding to $^{11}$Li the total angular
momenta are $J = 1/2, 3/2$ and $5/2$. They are shown in fig.~9 as
functions of the spin splitting parameter $\gamma_s$. The ground state
energy is also shown for comparison and as expected seen to be almost
independent of the spin splitting. In contrast the energies of the
excited states vary with $\gamma_s$. An increase is seen for
$J=1/2,3/2$ and a decrease for $J=5/2$. Crudely speaking the
$1^-$-state consists of one neutron in an s-state and the other
neutron in a p-state relative to the core.  By selecting the lowest
lying of the (spin split) s-states the energy of the three-body state
could therefore be expected to decrease with the energy of one of the
s-states in the neutron-core subsystem, i.e.~as soon as $\gamma_s \ne
0$. However, the $1^-$ states have admixtures, determined by a complex
interplay of several factors, of spin-split s-waves different from the
ground state.  Due to the antisymmetry requirement it is impossible to
construct a wavefunction with given total J and one neutron in the
lowest neutron-core spin split s-state and the other neutron in a
p-state. Both virtual s-states are necessarily present.  The J=5/2
state has an excess of neutron-core states with angular momentum
$s_{nc} = s_c+1/2 = 2$, the J=1/2 has analogously an excess of $s_{nc}
= s_c-1/2 = 1$ and the J=3/2 state is somewhere in between.

{}From the interaction in eq.~(\ref{e15}) we see that a negative value
of $V_s\gamma_s$ implies that the energy of the neutron-core state
with angular momentum $s_c+1/2$ is lower than that of $s_c-1/2$. In
our specific case of $s_c=3/2$ this corresponds for positive
$\gamma_s$ ($V_s <0 $) to a lower energy for the state of 2 than for
that of 1.  Thus the energy of the state with the largest content of 2
must exhibit the fastest decrease with $\gamma_s$.  Therefore with
increasing $\gamma_s$ the J=5/2 state necessarily becomes lowest in
energy followed by the J=3/2 state and with J=1/2 as the highest.  The
details of the resulting energies are seen in fig.~9. An increase of
the ground state energy would bring the excited spectrum into the
continuum, but a similar structure would remain.

\section{Summary and conclusion}
We discuss in this paper the effects of a finite core spin on the
structure of two-neutron halo nuclei. In the previous two papers in
this series we established the general connections between size and
binding energy for loosely bound three-body systems. The possible
asymptotic (spatially extended and small energy) structures were
characterized by the interactions between the constituent binary
subsystems. The largest of the ``particles'' called the core is
assumed to be inert and spinless and the corresponding degrees of
freedom are frozen.  When the core has a finite spin the two-body
interaction between the core and another particle (typically a
neutron) depends on the total spin of the two-body system. More
complicated spin structures therefore seems to be possible even when
the core still remains inert. The present paper, number III in the
series, is devoted to study effects of such spin couplings on the
three-body system.

The recently developed method to solve the Faddeev equations is well
suited for our purpose. We therefore first described the necessary
ingredience of our procedure. The definitions and notation are then
established in general and we proceed to consider the crucial two-body
interactions. The neutron-neutron potential is needed in both relative
s- and p-waves and since the spin structure is under investigation we
also have to distinguish between the singlet and triplet states with
the different total angular momenta. Low energy properties are still
expected to be decisive and we construct therefore an interaction
which reproduces the four phenomenologically extracted s- and p-wave
scattering lengths and in addition the s-wave effective range. This is
achieved by use of spin-spin, spin-orbit and tensor terms with a
common gaussian radial dependence. The resulting effective low energy
neutron-neutron interaction might also be useful in other connections.

The unknown neutron-core effective potential is also parametrized with
a gaussian radial shape. The minimum spin dependence include a
spin-orbit interaction and also the central interaction is allowed to
be different in each of the relative orbital angular momentum
channels.  We furthermore for s-waves include a term to distinguish
between the possible couplings of the core- and neutron spins and we
choose the scalar product of these spins.  The resulting interaction
is now flexible enough to study the influence on the three-body system
of the spin splitting and energy dependence of the resonances in the
binary subsystems.

Equipped with an efficient method and the input properly parametrized
we then explicitly write down the angular eigenvalue equations when
only s-waves are included. These states are often the most essential
components in the solutions. We find analytically both small and large
distance behavior of the angular eigenvalues. The conditions for the
Efimov effect are simultaneously obtained and investigated in terms of
scattering lengths for the two different spin couplings of the
neutron-core subsystem. The extreme Efimov effect, where infinitely
many bound states are present in the three-body system only occurs
when both scattering lengths are infinitely large.  Already the first
of these excited states is very extended in space compared to the
ranges of the potentials and the conditions for its presence is
therefore interesting even when no other excited state exists. These
conditions may conveniently be formulated in terms of bound state and
virtual s-state energies of the neutron-core subsystem. Both the two
spin couplings of $s_c \pm 1/2$ must correspond to energies between
about $-$2~keV and 5~meV. Since the typical spin splitting of such
s-states is fractions of an MeV, the occurrence of the first Efimov
state is extremely unlikely for a halo nucleus with a finite core
spin.

The spatial structure of the three-body ground state turns out to be
essentially independent of the spin splitting of the virtual s-state.
The reason is that either none or both s-states are needed in one
specific combination, if a total spin zero state has to be formed from
the spins of the two neutrons. Any other combination has the wrong
parity. On the other hand the spatial extension depends on the amount
of s-state admixture as noted in previous publications and easily
understood from the general behavior of states with higher angular
momenta or higher hyperspherical quantum numbers.

The energy of the three-body ground state depends sensitively on the
energies of the (spin split) virtual neutron-core s-states, which
constitute the components of the total system. The connection between
the properties of the neutron-core system and the total three-body
system depends strongly on the spin splitting.  This may be
exemplified by $^{10}$Li and $^{11}$Li, where a given s-state content
in the $^{11}$Li-wavefunction requires a specific combination the two
virtual s-states in $^{10}$Li. If one is low lying the other energy
must be high in order to compensate and lead to the observed total
three-body binding energy. Thus, both virtual s-states in $^{10}$Li
must be known, if the $^{11}$Li-binding energy should be accounted for
unambiguously.

The neutron-core s-state admixture in $^{11}$Li is, as recently
suggested, possible to extract from the measured momentum
distributions in nuclear break-up reactions. In the sudden
approximation, where one of the neutrons is removed instantaneously,
there is a finite probability of finding the other neutron in the
lowest lying virtual s-state. Due to the strong attraction this part
of the wavefunction leads to a narrow observable momentum
distribution.

The most obvious spin dependent observable is the magnetic dipole
moment, which for $^{11}$Li is found to be fairly close to the core
value measured for $^9$Li. The results of our strict three-body
calculations are that the core-value is reproduced independent of spin
splitting and s-state admixture. The reason is that the spins of the
two neutrons always couple to zero even when other possibilities are
allowed.

The discrete or continuous spectrum of excited states of
$1^-$character on top of the ground state is of special interest for
halo nuclei. One of the main components in the three-body wavefunction
consists of one neutron in an s-state and the other neutron in a
p-state relative to the core. Coupling of these angular momenta to 1
unit of $\hbar$ corresponds to an electric dipole excitation.  More
than one electric dipole excitation corresponding to different angular
momenta is possible for finite core spin. The related hyperfine
splitting of the E1-excitation leads to predictable structure in the
continuum.  Analyses of for example Coulomb dissociation cross
sections may elucidate the structure of this $1^-$ excitation
spectrum.

The excitation energy of such states may for a given total energy
depend strongly on the spin splitting of the virtual s-states. The
reason is that the excited states are composed of mixtures of spin
split s-states different from that of the ground state. On average the
lowest s-state is preferred and the statistically weighted excitation
energies therefore decrease when the energy of the lowest virtual
s-state approaches zero. The absolute value of this average depends
on the energy of the p-state resonance.

In conclusion, we have investigated effects of finite core spin in
halo nuclei. We found essentially unchanged spatial structure and
magnetic dipole moments. Efimov states seems to be extremely unlikely
in halo nuclei with finite core spin. The electric dipole excitations
may depend rather strongly on the relative positions of the resonances
and virtual states in the neutron-core subsystem. The spin splitting
of the s-states in the neutron-core system is decisive to understand
the proper relation between the total three-body system and its
subsystems.  Our numerical examples concentrated on
$^{11}$Li (neutron-neutron-$^9$Li) and $^{19}$B
(neutron-neutron-$^{17}$B).  These examples are used for general
illustration, but they are useful in their own right as model
computations for these nuclei.

{\bf Acknowledgments} One of us (DVF) acknowledges the support from
the Danish Research Council and from INFN, Trento, Italy.

\newpage
\begin{table}[h]
\centerline{Table 1}
\begin{center} Diagonal contributions to the magnetic moment from two-body
s- and p-states in the neutron-neutron Jacobi coordinate system. The
total and core angular momenta and parities are given by
$J^\pi~=~3/2^-$~and~$s_c^\pi~=~3/2^-$. The non-vanishing off diagonal
matrix elements are shown at the bottom of the table.
\end{center}
\center{\begin{tabular}{|c|ccccc|c|c|c|c|c|}
\hline
$no$ & $l_{nn}$ & $l_c$ & $L$ & $s_{nn}$ & $S$ & $\langle({\bf
s}_{nn})_z\rangle$ & $\langle({\bf s}_c)_z\rangle$ & $\langle({\bf
l}_c)_z\rangle$ & $\mu(^{11}{\rm Li})$ & $\mu(^{19}{\rm B})$ \\
\hline
1 & 0 & 0 & 0 & 0 & 3/2 & 0     & 3/2     & 0    & 3.4391=$\mu_c$ &
3.7928=$\mu_c$
\\
2 & 1 & 1 & 0 & 1 & 3/2 & 2/5   & 11/10   & 0    & 0.99 & 1.25 \\
3 & 1 & 1 & 1 & 1 & 1/2 &$-$1/3   & 5/6     & 1/2  & 3.46 & 3.65 \\
4 & 1 & 1 & 1 & 1 & 3/2 & 22/75 & 121/150 & 1/5  & 0.84 & 1.02 \\
5 & 1 & 1 & 1 & 1 & 5/2 & 21/25 & 63/50   &$-$3/10 &$-$0.49 & $-$0.19 \\
6 & 1 & 1 & 2 & 1 & 1/2 & 1/5   &$-$1/2     & 9/10 &$-$1.42 & $-$1.56 \\
7 & 1 & 1 & 2 & 1 & 3/2 & 2/25  & 11/50   & 3/5  & 0.53 & 0.57 \\
8 & 1 & 1 & 2 & 1 & 5/2 & 13/25 & 39/50   & 1/10 &$-$0.15 & 0.04 \\
\hline
&  & $\langle 2|$ & ${\bf \mu}$ & $|4 \rangle$ & & 0 & 0 & $\sqrt{2/5}$ & 0.34
& 0.33\\
&  & $\langle 3|$ & ${\bf \mu}$ & $|4 \rangle$ & & $-\sqrt{2/3}$ &
$-\sqrt{2/3}$ & 0 & 0.72 & 0.61 \\
&  & $\langle 3|$ & ${\bf \mu}$ & $|6 \rangle$ & & 0 & 0 & $-\sqrt{1/20}$ &
$-$0.12 & $-$0.12 \\
&  & $\langle 4|$ & ${\bf \mu}$ & $|5 \rangle$ & & $-\sqrt{54}/25$ &
$-\sqrt{54}/25$ & 0 & 0.45 & 0.38 \\
&  & $\langle 4|$ & ${\bf \mu}$ & $|7 \rangle$ & & 0 & 0 & 2/5 & 12/55 & 4/19
\\
&  & $\langle 5|$ & ${\bf \mu}$ & $|8 \rangle$ & & 0 & 0 & $\sqrt{21}$/10 &
0.25 & 0.24 \\
&  & $\langle 6|$ & ${\bf \mu}$ & $|7 \rangle$ & & $-\sqrt{2/5}$ &
$-\sqrt{2/5}$ & 0 & 0.97 & 0.82 \\
&  & $\langle 7|$ & ${\bf \mu}$ & $|8 \rangle$ & & $-3\sqrt{14}/25$ &
$-3\sqrt{14}/25$ & 0 & 0.69 & 0.58 \\
\hline
\end{tabular}}
\end{table}
\newpage

\begin{table}[h]
\centerline{Table 2}
\begin{center}
Diagonal contributions to the magnetic moment from two-body s- and
p-states in the neutron-neutron Jacobi coordinate system. The total
and core angular momenta and parities are given by
$J^\pi~=~3/2^-$~and~$s_c^\pi~=~1/2^-$. The non-vanishing off diagonal
matrix elements are shown at the bottom of the table. The gyromagnetic
factors of the core are respectively $g_c=3.4391/1.5$ and $g_c=3.7928/1.5$.
\end{center}
\center{\begin{tabular}{|c|ccccc|c|c|c|c|c|}
\hline
$no$ & $l_{nn}$ & $l_c$ & $L$ & $s_{nn}$ & $S$ & $\langle({\bf
s}_{nn})_z\rangle$ & $\langle({\bf s}_c)_z\rangle$ & $\langle({\bf
l}_c)_z\rangle$ & $\mu(^{11}{\rm Li})$ & $\mu(^{19}{\rm B})$ \\
\hline
1 & 0 & 2 & 2 & 0 & 1/2 & 0     & $-$3/10   & 9/5  &  0.29 & 0.19 \\
2 & 2 & 0 & 2 & 0 & 1/2 & 0     & $-$3/10   & 0    & $-$0.69 & $-$0.76 \\
3 & 1 & 1 & 1 & 1 & 1/2 & 2/3   & $-$1/6    & 1/2  & $-$2.66  & $-$2.71 \\
4 & 1 & 1 & 2 & 1 & 1/2 &$-$2/5   & 1/10    & 9/10 &  2.25  & 2.26 \\
5 & 1 & 1 & 0 & 1 & 3/2 & 1     & 1/2     & 0    & $-$2.68  & $-$2.56 \\
6 & 1 & 1 & 1 & 1 & 3/2 & 11/15 & 11/30   & 1/5  & $-$1.86  & $-$1.77 \\
7 & 1 & 1 & 2 & 1 & 3/2 & 1/5   & 1/10    & 3/5  & $-$0.21 & $-$0.20 \\
\hline
&  & $\langle 3|$ & ${\bf \mu}$ & $|4 \rangle$ & & 0 & 0 & $1/\sqrt{20}$ & 0.12
& 0.12 \\
&  & $\langle 3|$ & ${\bf \mu}$ & $|6 \rangle$ & & $-2/\sqrt{45}$ &
$-2/\sqrt{45}$ & 0 & 0.46 & 0.39\\
&  & $\langle 4|$ & ${\bf \mu}$ & $|7 \rangle$ & & $-$2/5 & $-$2/5 & 0  & 0.61
& 0.52 \\
&  & $\langle 5|$ & ${\bf \mu}$ & $|6 \rangle$ & & 0 & 0 & $\sqrt{2/5}$ & 0.34
& 0.33 \\
&  & $\langle 6|$ & ${\bf \mu}$ & $|7 \rangle$ & & 0 & 0 & 2/5 & 12/55 & 4/19
\\
\hline
\end{tabular}}
\end{table}

\newpage

\noindent{\Large\bf{Figure Captions}}
\begin{list}{}{\setlength{\leftmargin}{18mm}
\setlength{\labelwidth}{16mm}
\setlength{\labelsep}{2mm}}

\item[Figure 1\hfill] The relation between the virtual $s_{1/2}$-state and
the $p_{1/2}$-resonance for a total binding energy of 0.30~MeV.  The
curve is obtained by varying $V_s$ and $V_{so}^{(s)}$ while keeping
$V_p = -7.8~MeV$ and $\gamma_s = 0$.  The square, the diamond and the
dashed lines (corresponding to the asymptotic values) indicate points
of interest in the following figures.

\item[Figure 2\hfill]  The relation between the two virtual s-states
(total neutron-core spin $S_{nc}=1,2$) for a total binding energy of
0.30~MeV for $^{11}$Li. The curves are obtained by varying $V_s$ and
$\gamma_s$ while the other interaction parameters are defined by the
points indicated in figure 1 by the vertical line (circles) and the
square (squares).

\item[Figure 3\hfill] The total energy of $^{19}$B as function of the
energy E($s_{1/2}$) of the second virtual s-state. The energy of the
$d_{5/2}$-resonance in the neutron-core system is given by
E($d_{5/2}$)=xE($s_{1/2}$) where x is constant along the curves. The
parameters in the calculations are $V_{so}^{(d)}=-10~MeV$, and
$\gamma_s = 0$ while $V_s$ and $V_d$ are varied.

\item[Figure 4\hfill] The same as fig.~2 for $^{19}$B. The parameters
correspond to the black triangles in fig.3 with the total energy equal
to 0.45~MeV and 0.75~MeV. (The curves are obtained by varying
$\gamma_s$.)

\item[Figure 5\hfill] Large distance behavior of the angular eigenvalue
$\tilde\lambda$ as function of $\rho$ for $^{11}$Li. The scattering
lengths (marked by the vertical arrows) are $a_- = 10^2$~fm and $a_+ =
10^7$~fm for the dashed line and $a_+ = 10^2$~fm and $a_- = 10^7$~fm
for the solid line. The horizontal dashed lines correspond to the
various limiting values described in the text.

\item[Figure 6\hfill]  The angular eigenvalues as function of the radial
coordinate $\rho$ for interaction parameters corresponding to
the diamond on fig.~1, i.e~$^{11}$Li. Only antisymmetric states in the
neutron-neutron exchange are exhibited.

\item[Figure 7\hfill] The root mean square radius of $^{11}$Li and
$^{19}$B as functions of the energy of one of the virtual s-states for
interaction parameters corresponding to the curves in figs.~2 and~4. The
squares, circles and triangles have the same meaning as in those figures.

\item[Figure 8\hfill]   The root mean square radii of $^{11}$Li (open
circles) and $^{19}$B (black circles) as functions of the s-state
admixture in the neutron-core channel. The core radii are
$R_{RMS}(^{9}{\rm Li})$ = 2.32~fm and $R_{RMS}(^{17}{\rm B})$ = 2.65~fm
respectively.  The interaction parameters correspond to the curve in
fig.~1 and the horizontal dashed line in fig~3.

\item[Figure 9\hfill] The ground state and the excited $1^-$-levels
as functions of the spin splitting parameter $\gamma_s$. The
interaction parameters are $V_s = -8.5$, $V_p = -7.8$ and
$V_{so}^{(p)}=34.3~MeV$.

\end{list}

\end{document}